# Performance and Complexity Analysis of Bi-directional Recurrent Neural Network Models vs. Volterra Nonlinear Equalizers in Digital Coherent Systems

Stavros Deligiannidis, Charis Mesaritakis, Adonis Bogris, *Senior Member OSA*

*Abstract*—We investigate the complexity and performance of recurrent neural network (RNN) models as post-processing units for the compensation of fibre nonlinearities in digital coherent systems carrying polarization multiplexed 16-QAM and 32-QAM signals. We evaluate three bi-directional RNN models, namely the bi-LSTM, bi-GRU and bi-Vanilla-RNN and show that all of them are promising nonlinearity compensators especially in dispersion unmanaged systems. As far as inference is concerned, our simulations show that the three models provide similar compensation performance, therefore, in real-life systems, the simplest scheme based on Vanilla-RNN units should be preferred. We compare bi-Vanilla-RNN in its many-to-many form with Volterra nonlinear equalizers and exhibit its superiority both in terms of performance and complexity, thus highlighting that RNN processing is a very promising pathway for the upgrade of long-haul optical communication systems utilizing coherent detection.

## I. INTRODUCTION

Optical communication links are the main highways for the exchange of trillions of data around the globe every day. The ever-increasing deployment of next generation mobile communication systems and cloud/edge computing infrastructures pushes the limits related to the required bandwidth capacity of optical communication links and enhances the need for ultra- high speed long-haul transmission. The optical communication society has launched a variety of techniques so as to deal with this demand. Current solutions involve a combination of advanced modulation formats [1] with the use of space division multiplexing [2] and bandwidth extension towards other bands such as O-band [3]. Regardless the capacity enhancement method that is adopted, the major limitation factor of capacity will eventually be the nonlinear Shannon capacity limit of transmitted information [4, 5]. In long-haul high bandwidth optical networks, this limit is mainly attributed to Kerr-induced fibre nonlinearities and their interaction with amplified spontaneous emission noise from cascaded optical amplifiers [5].

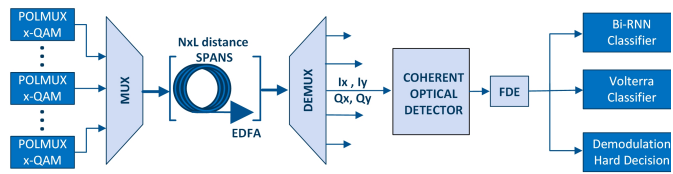

Fig. 1. The simulated transmission link.

Principal techniques that deal with nonlinearity compensation are mid-span optical phase conjugation (OPC) [4, 6], digital back-propagation (DBP) [7], and inverse-Volterra series-transfer function (IVSTF) [8]. OPC is a purely analog and thus ultra-fast technique, however it needs extra hardware such as low-noise and broadband wavelength converters and poses limitations in the transmission link topology as phase conjugation must take place at specific points with respect to the total link of a lightpath [6]. DBP is one of the most suitable post-processing techniques, proper for the treatment of both linear and nonlinear deterministic effects, since it emulates almost perfectly fiber channel through split-step Fourier with the exception of signal-noise interactions and polarization mode dispersion; however, its real-life implementation still remains impractical due to its high complexity especially when DBP is used to emulate and thus mitigate the effects of a multi-channel transmission scenario [9]. IVSTF is a less complex variant compared to DBP, however in principle it is more appropriate for mitigating intra-channel nonlinearity [8, 10]. Nonlinear Fourier transform is a promising alternative currently being investigated in the community [11, 12]. Lately, there exists an increasing interest in the investigation of machine learning techniques for the mitigation of transmission impairments [13]. Different paradigms based on artificial neural networks (ANNs) [14], convolutional neural networks (CNNs) [15], recurrent neural networks (RNNs) [16] are among the techniques that have been successfully applied mostly in intensity

Manuscript received XXXX.
This work has been partially funded by the H2020 project NEoteRIC (871330).
Stavros Deligiannidis and Adonis Bogris are with the Department of Informatics and Computer Engineering, University of West Attica, Aghiou Spiridonos, Egaleo, 12243, Athens, Greece (e-mail: sdeligiannid@uniwa.gr, abogris@uniwa.gr).
Charis Mesaritakis is with the Department of Information & Communication Systems Engineering, University of the Aegean, 2 Palama & Gorgyras St., 83200, Karlovassi Samos, Greece (e-mail: cmesar@aegean.gr).

TABLE I
NUMERICAL MODEL PARAMETERS

| Symbol | Parameter | Value |
| --- | --- | --- |
| G | gain of amplifier | 10dB @ 50 km or 16 dB @ 80 km |
| a | attenuation | 0.2dB/km |
| $\beta_2$ | second order dispersion | -21ps$^2$/km |
| $\gamma$ | fibre nonlinear coefficient | 1.3 W$^{-1}$km$^{-1}$ |
| R | symbol rate / channel | 25 Gbaud, 32 Gbaud, 64 Gbaud |
| M | modulation format | Dual-polarization 32-QAM or 16-QAM |
| L | span distance | 50 km or 80 km |
| $\Delta f$ | Channel spacing | 35 GHz or 75 GHz @ 64 Gbaud |

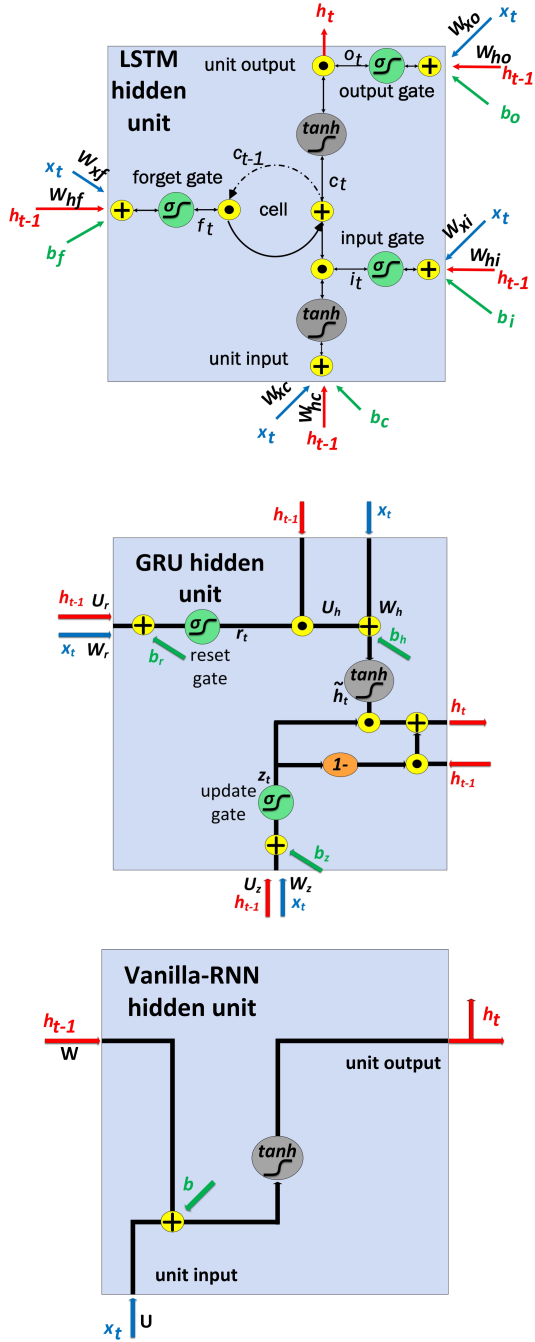

Fig. 2. Conceptual illustration of the LSTM, GRU and Vanilla-RNN units

modulation/direct detection systems (IM/DD) and in orthogonal frequency division multiplexing (OFDM) [13].
Very recently, we proposed for the first time the utilization of a Long Short-Term Memory (LSTM) network, which is a well-known efficient RNN model for the compensation of fibre nonlinearities in digital coherent systems for multi-channel polarization multiplexed 16-QAM systems [17]. A detailed analysis regarding the effect of LSTM model parameters and channel memory was conducted in order to reveal the properties of LSTM based receiver with respect to performance and complexity in comparison to Digital Back Propagation (DBP), proving its mitigation superiority in inter-channel transmission effects. In [18] we extended the analysis by considering two additional RNN models that are in principle less complex than LSTM, that is Gate Recurrent Unit (GRU) and Vanilla RNN in order to investigate the potential of adopting bidirectional RNN models in next generation digital coherent optical communication systems at moderate complexity. Very recently, the work of [19] confirmed experimentally the performance of the proposed RNNs in single channel operation and M. Schaedler et al. proposed a recurrent neural network soft-demapper in 800G-DWDM-600 km [20] and a soft Deep Neural Network for short reach optical communications [21].

In this work, we provide a comparative analysis of the three RNN models, namely LSTM, GRU and Vanilla RNN in WDM systems with small channel spacing and high-order modulation formats. Moreover, we compare the bi-RNN models with a 3$^{rd}$ order Volterra nonlinear equalizer, as a reference technique with the ability to treat complex time-dispersive nonlinear effects [22, 23], both in terms of performance and complexity. We clearly show that with the use of many-to-many training, we drastically reduce the complexity of RNN models vs. Volterra whilst keeping their performance superiority in terms of the bit-error rate (BER) of the decoded signal. The results of the present work clearly reveal the potential of bi-RNN nonlinear signal processing in next generation optical coherent communication systems as efficient and low complexity post processors. Next section describes system modelling in detail.

## II. SYSTEM MODELING

The system depicted in Fig. 1 simulates the up to 1000 km fiber transmission, numerically simulated with the integration of Nonlinear Schrodinger equation (NLSE). Fibre propagation was modelled based on Manakov's equations [24] using split-step Fourier method.

$$\frac{\partial E_{x,y}}{\partial z} = -\frac{a}{2}E_x + \frac{j\beta_2}{2}\frac{\partial^2 E_{x,y}}{\partial t^2} - j\gamma\frac{8}{9}\left(|E_x|^2 + |E_y|^2\right)E_{x,y} \quad (1)$$

We consider 9-channel dense wavelength division multiplexing (DWDM) transmission with polarization multiplexing. In (1), $E_{x,y}$ contain the overall field of the nine co-propagating wavelengths including their frequency detuning with respect to the central wavelength, thus taking into account all inter-channel nonlinear effects. We calculate the bit error rate (BER) in all scenarios studied and we aim at achieving BER < 3.8x10$^{-3}$ as dictated by the hard-decision forward error correction (HD-FEC) which adds a 7% overhead. We test 16-QAM 32 Gbaud per polarization which corresponds to 256 Gb/s line rate (238 Gb/s net data rate) per wavelength and then we extend the analysis to 32-QAM at 25 Gbaud (250 Gb/s line rate and 232.5 Gb/s net data rate) and 16QAM at 64 Gbaud (512 Gb/s line rate, 476 Gb/s net data rate) so as to have a variety of modulation formats and baud rates. Net data rate is derived considering that only FEC overhead is taken into account. Lumped amplification was used with span length equal to 80 km or 50 km and noise figure equal to 5 dB. All parameters are summarized and provided in Table I. The optical receiver depicted in fig. 1 consists of an optical hybrid, balanced photodetectors, low-pass electrical filters with cut-off frequency matched to the baud rate. We assumed ideal carrier phase and frequency estimation as well as polarization demultiplexing as we want to solely focus on nonlinear impairments and their mitigation. Prior to any post-processing or demodulation, we first sample the signal and then perform chromatic dispersion compensation with the use of an ideal frequency domain equalizer (FDE). The simulations were conducted with pulse shaping. They incorporate root-raised-cosine (RRC) shaping, with a roll-off factor of 0.2. In this paper we numerically investigate the efficiency of three types of Bidirectional Recurrent Neural Networks (bi-RNN) in compensating fibre communication systems, namely LSTM, GRU and Vanilla-RNN as already reported in the introduction. Fig. 2 illustrates the RNN units that we use whilst Eq. (2)-(4) indicate how the output $h_t$ is calculated in each case.

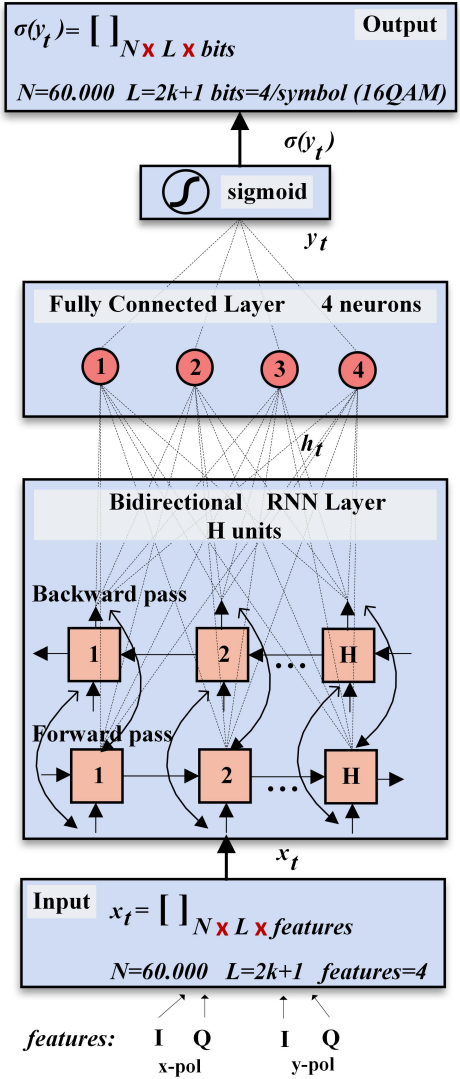

Fig. 3. Bidirectional-RNN model architecture in the case of 16-QAM

$$i_t = \sigma(W_{xi}x_t + W_{hi}h_{t-1} + b_i)$$
$$f_t = \sigma(W_{xf}x_t + W_{hf}h_{t-1} + b_f)$$
$$o_t = \sigma(W_{xo}x_t + W_{ho}h_{t-1} + b_o)$$
$$c_t = f_t * c_{t-1} + i_t * tanh(W_{xc}x_t + W_{hc}h_{t-1} + b_c)$$
$$h_t = o_t * tanh(c_t) \qquad (2, \text{LSTM})$$

$$z_t = \sigma(W_z\, x_t + U_z h_{t-1} + b_z)$$
$$r_t = \sigma(W_r\, x_t + U_r h_{t-1} + b_r)$$
$$\tilde{h} = tanh[W_h\, x_t + U_h(r_t * h_{t-1}) + b_h]$$
$$h_t = (1 - z_t) * h_{t-1} + z_t * \tilde{h} \qquad (3, \text{GRU})$$

$$h_t = tanh\,(W\, h_{t-1} + U\, x_t + b\,) \qquad (4, \text{Vanilla} - \text{RNN})$$

where $W$ matrices contain the weights of connection: $f$, $i$, $o$ and $c$ stands for forget, input, output gate and cell state respectively in the case of (2), $W$ and $U$ matrices contain the weights of connection: $z$, $r$, $\tilde{h}$ stands for update, reset gate and candidate activation vector in the case of (3, 4), $x_t$, $h_t$, $h_{t-1}$ are input, hidden output, previous hidden output and $b$ are bias vectors. The * operator denotes the element wise product, $\sigma$ is the logistic sigmoid function and $tanh$ is the hyperbolic tangent activation function.

The sequential neural model is demonstrated in Fig. 3. The input $x_t$ is the distorted symbol sequence which has the following form $x_{t,L}=[x_{t-k},…,x_{t-1}, x_t, x_{t+1},…,x_{t+k}]$, where $L$ stands for the overall length of the word which is equal to $L=2k+1$. Thus, for the symbol at time $t$ we also launch $k$ preceding and $k$ succeeding symbols so as to track intersymbol dependencies. The length of $L$ depends on the foreseen channel memory strictly related to accumulated chromatic dispersion. Each symbol in each window contains four values/features (I and Q for both polarizations) as the input Xx-pol and Xy-pol feeding the Bi-RNN layer of $L$ hidden units. In order to calculate bit BER we drive the RNN network output $h_t$ to a Fully Connected Layer of 4 neurons in the case of 16-QAM or 5 neurons in the case of 32-QAM and then $y_{t,L}=[y_{t-k},…,y_{t-1}, y_t, y_{t+1},…,y_{t+k}]$ to a sigmoid layer that carries out the bit-wise estimation $\sigma(y_t)$ for all the symbols at the output. The bit-wise approach [20] is slightly different compared to the symbol-wise approach of [17] and marginally improves the complexity of the model without sacrificing the BER performance.

We train the model using many input and many output symbols, (many to many approach) which predicts simultaneously the same number of symbols $y_t$ as those of input $x_t$ [25]. The RNN models are built, trained and evaluated in Keras with Tensorflow 2.3 GPU backend. In the Keras model, binary crossentropy is chosen as a loss function and Adam as the optimizer for the BER measurement with the parameters

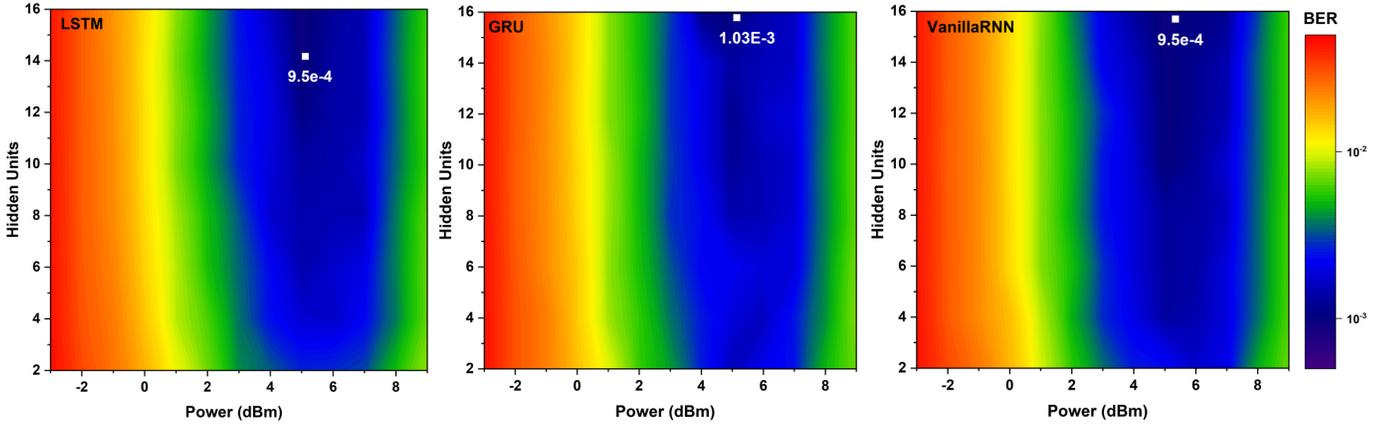

Fig. 4. BER as a function of total launched optical power and number of hidden units for the bi- LSTM (a), bi-GRU (b) and bi-Vanilla-RNN (c) for a 960 km (12 spans of 80 km each) optical transmission link, dispersion of -21 ps$^2$/km carrying 16-QAM 32 Gbaud signals. The system based only on FDE exhibited minimum BER equal to 4x10$^{-3}$

appearing in [26]. We consider 40.000 symbols for training, 20.000 for validation and 60.000 for testing with unknown data. The training stage is executed with batches of 512 words of symbols for optimum balance between memory allocation size and execution time. The maximum forward and backward

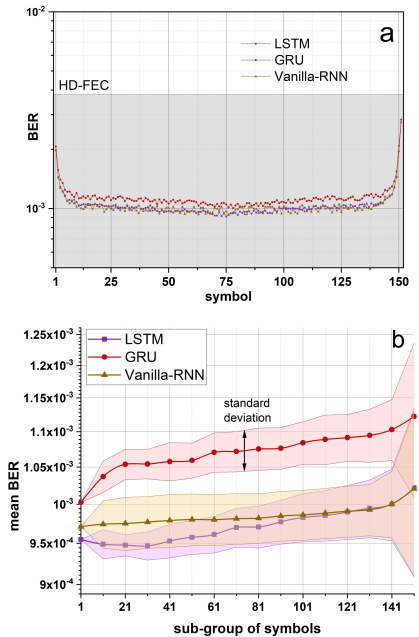

Fig. 5. a) The value of the BER for each of the 151 symbols of the word, for all RNN models. 100 symbols have approximately the equal minimum BER value, while a total of 100 symbols (green zone) exhibit BER < 10$^{-3}$. b) mean BER and its standard deviation calculated in a sub-group of symbols in the word of 151 symbols. Standard deviation explodes when more than 140 symbols are simultaneously detected

passes of all training sequences (epochs) are chosen to be 300. To avoid overfitting during training we use "early stopping" when validation loss does not decrease for 20 successive epochs. The RNN processor was applied to the central WDM channel that is in principle the most heavily impaired one and requires only one sample per symbol.

The third-order Volterra nonlinear equalizer used as an alternative to RNN post-processor was simulated as follows:

The equalized output $y(n)$ of the $x(n)$ sampled signal from a 3$^{rd}$ order Volterra series can be described according to (5).

$$y(n) = \sum_{i_1=-k_1}^{k_1} w_{i_1} x(n+i_1)$$

$$+ \sum_{i_1=-k_2}^{k_2} \sum_{i_2=i_1}^{k_2} w_{i_1,i_2} x(n+i_1) x(n+i_2)$$

$$+ \sum_{i_1=-k_3}^{k_3} \sum_{i_2=i_1}^{k_3} \sum_{i_3=i_2}^{k_3} w_{i_1,i_2,i_3} x(n+i_1) x(n+i_2) x(n+i_3) \quad (5)$$

where $k_j = (L_j - 1)/2$, in which $L_j$ is the input length of the $j$-$th$ order. $w_{i_1,\ldots,i_j}$ is the weight for each sample of the $j$-$th$ order [23]. We train the Volterra model trying to minimize error using the least mean square algorithm. We consider 60.000 symbols for training and 60.000 for testing with unknown data. Each input symbol is a vector of four features (I and Q for both polarizations). The following section presents the results of our numerical analysis starting with the comparison between different bi-RNN models.

III. RESULTS AND DISCUSSION

A. Bi-RNN Models - Performance Comparison

First, we compare the three bi-RNN models in terms of BER performance. We test their ability to identify 16-QAM modulated symbol-series of 32 Gbaud, that have been transmitted along 960 km (80 km spans) at 1550nm. The values of the other propagation parameters are in accordance with Table 1. We train the different bi-RNNs with a symbol word of 151 symbols that approximates nonlinear channel memory depending on accumulated dispersion [17]. In fig. 4, it can be seen that all models perform adequately and equivalently as they improve BER by almost half an order of magnitude (BER ~ 1x10$^{-3}$) compared to a transmission system that uses only linear equalization. At this point, it must be mentioned that a transmission system that incorporates only linear equalization with the use of FDE, achieves BER=4x10$^{-3}$ at minimum for the specific transmission parameters. It can be seen that bi-LSTM offers best performance for a minimum of 14 units, whilst bi-GRU and bi-Vanilla-RNN for 16 units. It is interesting that all models perform equivalently, thus there is no substantial need to use demanding and complex bi-RNN models based on LSTM or GRU units. Bi-Vanilla-RNN topology with almost the same capacity, as dictated by the number of hidden units, but significantly simpler architecture can exhibit similar performance based on findings of fig. 4. It is well-known that LSTM models use the *cell, input, output and forget gate* structure [27] and GRU models use the *update gate* mechanism, in order to simultaneously prevent the vanishing gradient problem [28] and handle long-term or short-term dependencies [29], namely lags of unknown duration between important events in a time series. The intersymbol interference caused by fibre propagation, as a result of accumulated dispersion, determines channel memory, which remains constant as long as the propagation parameters remain unchanged. Once the network learns the fibre channel memory, it acquires knowledge of the position of the symbols in the sequence and there are no unexpected or time-varying long-term or short-term dependencies as for example exist in rapidly time-varying wireless communication channels. Therefore, the complex aforementioned mechanism of LSTM and GRU becomes unnecessary. Based on the aforementioned finding, the training becomes a process that can be easily undertaken by a bi-Vanilla-RNN model, provided that the symbol word length is long enough to embrace channel memory and nonlinear intersymbol interference. Regarding training efficiency, in all training simulations we carried out, we observed that all RNNs converge more or less to the same loss value but Vanilla-RNN needs about 30% more training epochs than LSTM and GRU, with the latter exhibiting the fastest training. Considering that the optical channel does not require frequent training, the training speed is not critical for the adoption or rejection of a specific RNN model. As already shown in [17], RNN models are robust against power fluctuations or even the modulation format of neighboring channels. Hence, the potential period of re-training relates to the time scales of polarization effects, that is a few ms which corresponds to the duration of hundreds of millions of symbols. Since the system will have been initially trained, re-training will need less symbols (A few thousand) and less time as its purpose will be fine tuning and not training from scratch. This can be accomplished even with the use of a parallel monitoring system which will feed the training unit with a small portion (< 1%) of the real time transmitted and inferred data so as to continuously adjust the weights and keep the system stable.

It must be mentioned that the BER value appearing in fig. 4 is the one obtained for the central symbol in $y_t$. Since our RNN models are trained using the many to many approach, one can simultaneously extract all symbols in the word. Fig. 5a shows the BER for all symbols in $y_t$. It becomes evident that the symbols at the center of the word enjoy optimal BER performance. Nevertheless, at least 100 symbols enjoy approximately the minimum BER value (~$1\times10^{-3}$ ±$0.5\times10^{-4}$) also obtained for the central symbol. When we approach the edges of the word, the performance gradually degrades, which confirms the need for training with enough neighboring preceding and succeeding symbols so that non-linear intersymbol interference effects are adequately captured. In the left edge, past symbols are not enough, whilst in the right edge, future symbols are those missing. Fig. 5b highlights this behavior in the form of mean BER and its standard deviation when BER is calculated for a portion of the symbols in the word (151 in our case). We formulate these sub-groups of symbols starting from the central symbol in the word and include the same number of preceding and succeeding symbols. For instance, when sub-group of symbols has a value of 21, then this sub-group contains the central symbol and 10 preceding and succeeding symbols. Based on fig. 5b, it becomes obvious that the mean BER remains constant even if 100 out of 150 symbols are simultaneously detected and its standard deviation explodes when subgroups contain more than 140 symbols. Thus, one can easily deduce that the simultaneous detection of 100 symbols in a word of 151 symbols is efficient and will not induce severe degradation of the BER with respect to the case where only the central symbol is decoded.

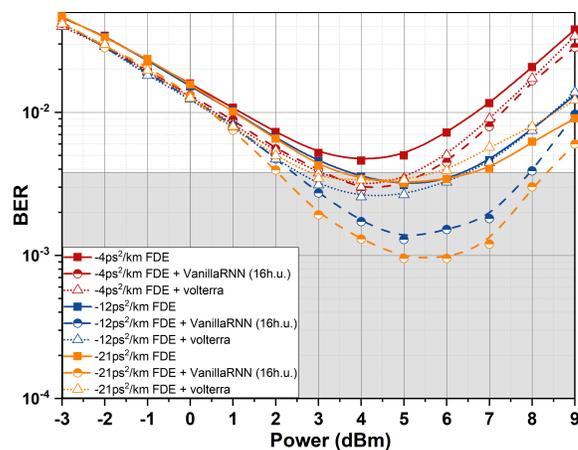

Fig. 6. BER as a function of optical launched optical power for dispersion of -4, -12 and -21 ps$^2$/km, with linear equalization (FDE) and with a bi-Vanilla-RNN of 16 hidden units or Volterra

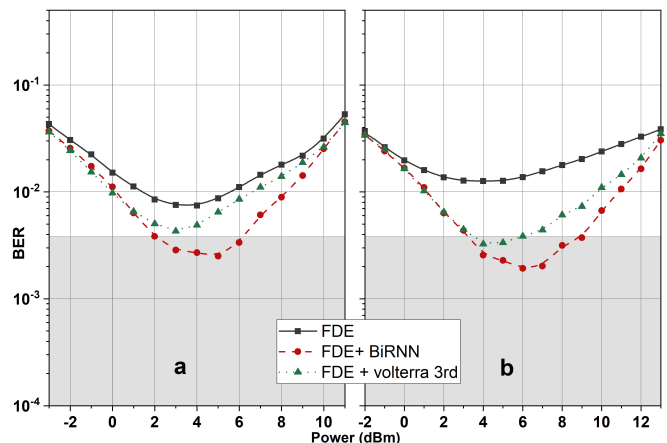

Fig. 7. BER as a function of launched optical power, a) 25GBaud 32-QAM (960 km, 50 km spacing) (b) 64 GBaud, 16-QAM (640 km, 80 km spacing)

In order to study the efficiency of bi-Vanilla-RNN model for different channel memory scenarios we carried out numerical simulations for -4, -12 and -21 ps$^2$/km second order dispersion values. Although in real systems a dispersion value different than that of a typical SMF (~-21 ps$^2$/km) is translated to the use of dispersion compensating modules or other types of fibres that differentiate other critical parameters of the link, here we assume that all the other parameters of the system are not affected in order to identify how the RNN-based equalizer behaves at different channel memories assuming that signal to noise ratio and nonlinearity are kept constant. We have adjusted the word length so as to effectively address memory effects of the highest studied dispersion. With careful training, this does not affect the performance of low memory channels [17]. Apart from FDE equalization, we also conducted numerical simulations using the Volterra nonlinear equalizer. For fair comparison of the two equalizers we use the same memory length ($L$=151 for RNN and $L_1$=151 for the 1st order of Volterra). As far as Volterra is concerned, we also considered $L_2$=51, $L_3$=11 for the 2nd and 3rd order kernels, two values which are sufficient for equalizing intra-channel nonlinearity impairments and offer the best performance at the least possible complexity as proved by extensive simulations we conducted. In fig. 6, one can see that FDE compensated transmission systems exhibit BER varying from to $4 \times 10^{-3}$ to $2 \times 10^{-3}$ with better values appearing in the highest accumulated dispersion values due to the fact that inter-channel nonlinear effects become less intense as dispersion increases. Bi-Vanilla-RNN compensation exhibits significantly better BER at larger accumulated dispersion. Almost identical BER behavior was verified for GRU and LSTM units as well and is related to the coherence time of the channel which is much longer than the symbol period as dispersion increases. Hence inter-channel effects become very slow and easily tracked by the Vanilla-RNN equalizer [17, 30]. Volterra equalizer slightly improves BER compared to linear equalization, however its performance is not improved at larger dispersions. Vanilla-RNN performs much better than Volterra as well as the former seems to adequately track inter-channel effects whilst the latter deals with intra-channel effects [31]

### B. Operation at different regimes

We tested the proposed bi-Vanilla-RNN model at different regimes by either increasing the order of modulation or the baudrate. Working with typical single mode fibers in the C-band (1550nm, -21 ps$^2$/km dispersion) we increased the order of the modulation format to 32-QAM and decreased the baud rate to 25 Gbaud. In fig. 7a, we observe that only bi-Vanilla-RNN detector is able to perform below the FEC limit, thus showing its systematic superiority in more complex modulation formats as well. It is stated out that GRU and LSTM models in this harsh environment did not exhibit better BER performance than Vanilla-RNN, thus their equivalence was once again proved. These results are not depicted. In order to align our study with the current state-of-the-art transmission systems we increased the baudrate/channel to 64 GBaud with 75 GHz channel spacing, we considered spans of 80 km for a total of 640 km transmission (fig. 7b). The simulation results, even for increased baudrate per channel (~ 500 Gb/s capacity per wavelength) verify that the proposed bi-Vanilla RNN equalizer is capable of mitigating nonlinearities in dispersive channels and once more show its superiority over nonlinear Volterra equalizers.

TABLE II
A. COMPUTATIONAL COMPLEXITY IN TERMS OF THE NUMBER OF REAL MULTIPLICATIONS – CENTRAL SYMBOL AS OUTPUT (16QAM)

| Equalizer | Memory Length(symbols) | Hidden units | Multiplications |
|---|---|---|---|
| Bi-LSTM | 151 | 16 | 386688 |
| Bi-GRU | 151 | 16 | 290048 |
| Bi-Vanilla-RNN | 151 | 16 | 96768 |
| **Volterra** | **151/51/11** | **-** | **14644** |

B. COMPUTATIONAL COMPLEXITY IN TERMS OF THE NUMBER OF REAL MULTIPLICATIONS – MANY SYMBOLS AT THE OUTPUT (16QAM)

| Equalizer | Memory Length | Hid. units | Output symbols | Multiplications per symbol |
|---|---|---|---|---|
| Bi-LSTM | 151 | 16 | 80 | 5074 |
| Bi-LSTM | 151 | 16 | 120 | 3382 |
| Bi-GRU | 151 | 16 | 80 | 3866 |
| Bi-GRU | 151 | 16 | 120 | 2577 |
| **Bi-Vanilla-RNN** | **151** | **16** | **80** | **1450** |
| **Bi-Vanilla-RNN** | **151** | **16** | **120** | **966** |
| Volterra | 151/51/11 | - | 1 | 14644 |

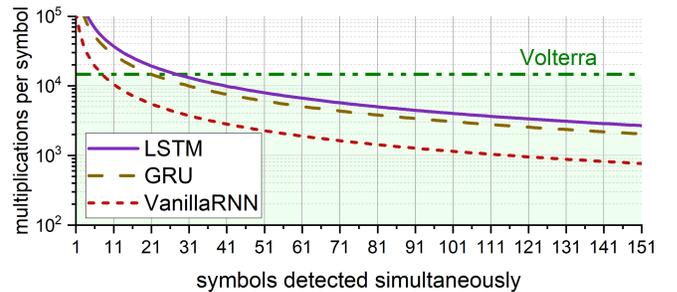

Fig. 8. Number of multiplications per symbol as a function of symbols detected simultaneously

### C. Complexity analysis

Finally, we investigated the receiver complexity focusing only on bi-RNN and Volterra units. FDE complexity has been analyzed in previous works [17]. In general, the overall bi-RNN equalizer complexity depends on the number of parameters (weights) that each network needs to calculate (see Eq. (2)-(5) and fig. 1), on the number of hidden units and the length of the input word. According to fig. 4, LSTM needs at least 14 units whilst GRU and Vanilla-RNN need 16 units to achieve optimal BER, thus the three models are more or less equivalent. Regarding the optimal word length, this does not depend on the selected model architecture, it is strictly related to channel memory. The only parameter differentiating the three models is the complexity of each unit. Based on (2)-(4), one can easily calculate the number of parameters for each model as follows:

$$bi - RNN_{param} = 2B[H(H + F) + H] + (2H + 1)b \quad (6)$$

where $B$=4,3,1 for LSTM, GRU and Vanilla-RNN respectively indicating the number of gates contained in each RNN unit (see fig. 2), $H$ the number of hidden units, $F$=4 the number of input features (see fig. 3) and $b$ the number of bits ($b$=4 in the case of 16-QAM). In order to directly compare bi-RNNs with Volterra nonlinear equalizer, it is necessary to calculate the computational complexity in terms of the number of real multiplications. For bi-RNNs we can calculate the number of multiplications, with the use of Eq. (7). We note that the bias is set to zero for all the bi-RNN models, thus it should not be taken into account in the complexity estimation.

$$bi-RNN_{mult} = 2B(FH + H^2)L + 2HbL \quad (7)$$

where $L$ is the length of the input symbol sequence.
The computational complexity $CC_{Volterra}$ of Volterra nonlinear equalizer of eq. (5) in terms of the number of real multiplications is given by Eq. (8) [23].

$$CC_{Volterra}(K,L) = \sum_{k=1}^{K} \frac{(L_k-1+k)!}{(k-1)!(L_k-1)!} \quad (8)$$

As Eq. 8 denotes, the number of multiplications explodes as the series order increases [23], this is why we have chosen $L_2$, $L_3$ to be smaller than $L_1$. In Table IIA, we compare the three models with Volterra in terms of the number of multiplications. The word length is kept at 151 symbols for all equalizers ($L_1$=151 for Volterra), whilst for the second and third order we have used $L_2$=51 and $L_3$=11 as explained in BER performance analysis. It can be seen that among the three bi-RNN models, bi-Vanilla-RNN is the least complex due to a lower $B$ parameter. In the case that RNN predicts/detects only the central symbol, the Volterra is far less complex than all RNN schemes, thus making their use not attractive. Nevertheless, this finding is misleading in the sense that the great advantage of the RNN training method, the "many to many" approach, is not fully exploited here. According to (7), one can easily ascertain that the number of multiplications per symbol in the many-to-many approach reduces with the number of symbols that can be simultaneously detected as shown in fig. 5. Table IIB and fig. 8 show the multiplications per symbol needed for the many-to-many approach which was used in this work when multiple symbols are simultaneously decoded. It is evident that taking into account that all RNN schemes offer simultaneous detection of multiple symbols, they are proved to be less complex than Volterra nonlinear equalizer that cannot support a similar mechanism for multiple symbol detection. The bi-Vanilla-RNN turns out to be the least complex when 120 output symbols of 151 are exported at expense of a slight degradation in the BER as depicted in fig. 5. Hence, the bi-RNN model can be almost 90% lighter than the Volterra nonlinear equalizer. It is worth noting that although there are a lot of studies for complexity reduction for both Volterra equalizers [23] and RNNs, like pruning [32] we decided to identify the complexity of the conventional models. Even if complexity reduction in the order of 70% is taken into account for Volterra based on [23], bi-Vanilla RNN is still preferrable in terms of complexity. Pruning is one of the subjects we will study for bi-RNN as well in the near future.

## IV. CONCLUSION

In this paper we numerically studied three bi-RNN models, (LSTM, GRU and Vanilla-RNN), as potential fibre nonlinearity compensators in high capacity digital coherent systems. At distances of 1000 km all models exhibited BER improvement of half an order of magnitude compared to systems utilizing exclusively linear equalization and three times lower compared to Volterra nonlinear equalizer. Their efficacy becomes stronger for dispersion unmanaged systems and their superiority over linear and Volterra nonlinear equalization was verified in many transmission scenarios. Among the three models, the bi-Vanilla-RNN exhibits the lowest complexity without lagging behind in BER performance. Finally, we estimated the complexity in terms of the number of multiplications per symbol in the "many to many" training and inference and proved that bi-RNN can be almost 90% less complex than a 3rd order Volterra nonlinear equalizer. Our work explicitly shows that bi-RNNs are a promising post processing method for mid-term deployment in coherent detection transmission systems. In the near future, we intend to study pruning techniques for further minimizing the complexity of RNN models.

**Stavros Deligiannidis** (holds a BSc in Physics, a MSc degree in Microelectronics and VLSI from the National and Kapodistrian University of Athens. Since 2010 he is with the Department of Computer Engineering of the Technological Educational Institute of Peloponesse, Greece where he serves as a Lecturer. He is currently pursuing his PhD degree at the University of West Attica in the field of novel signal processing techniques for optical communication systems under the supervision of Prof. Adonis Bogris. He has worked as a researcher in local and European projects. His current research interests include photonics, deep learning, digital signal processing, and parallel computing.

**Charis Mesaritakis** received his BS degree in Informatics by the department of Informatics & Telecommunications of the National & Kapodistrian University of Athens in 2004. From the same insitution he received the MSc in Microelectronics, whereas in 2011 he received his Ph.D degree on the field of quantum dot devices and systems for next generation optical networks, by the photonics technology & optical communication laboratory of the same institution. In 2012 he was awarded a European scholarship for post-doctoral studies (Marie Curie FP7-PEOPLE IEF) in the joint research facilities of Alcatel-Thales-Lucent in Paris-France, where he worked on intra-satellite communications. He has actively participated as research engineer/technical supervisor in more than 10 EU-funded research programs (FP6-FP7-H2020) targeting excellence in the field of broadband communications, cyber-physical security and photonic integration. He is currently an Associate Professor at the Department of Information & Communication Systems Engineering at the University of the Aegean, Greece. He is the author and co-author of more than 60 papers in highly cited peer reviewed international journals and conferences, two international book chapters, whereas he is serving as a regular reviewer for IEEE, OSA, AIP and Springer.

**Adonis Bogris** was born in Athens. He received the B.S. degree in informatics, the M.Sc. degree in telecommunications, and the Ph.D. degree from the National and Kapodistrian University of Athens, Athens, in 1997, 1999, and 2005, respectively. His doctoral thesis was on all-optical processing by means of fiber-based devices. He is currently a Professor at the Department of Informatics and Computer Engineering at the University of West Attica, Greece. He has authored or co-authored more than 150 articles published in international scientific journals and conference proceedings and he has participated in plethora of EU and national research projects. His current research interests include high-speed all-optical transmission systems and networks, nonlinear effects in optical fibers, all-optical signal processing and all-optical networking, nonlinear effects in lasers and photonic waveguides, mid-infrared photonic devices and cryptography at the physical layer. Dr. Bogris serves as a reviewer for the journals of the IEEE.